\documentclass[prl,twocolumn,showpacs,letterpaper,showpacs]{revtex4-1}
\usepackage{graphicx,amsmath,amssymb,amsfonts,latexsym,color,dcolumn,bm}

\newcommand{\beq}{\begin{equation}}
\newcommand{\eeq}{\end{equation}}
\newcommand{\bea}{\begin{eqnarray}}
\newcommand{\eea}{\end{eqnarray}}
\providecommand{\abs}[1]{\left\lvert#1\right\rvert}

\providecommand{\moy}[1]{\langle #1 \rangle}
\providecommand{\bra}[1]{\langle #1 \rvert}
\providecommand{\ket}[1]{\lvert #1 \rangle}
\providecommand{\braket}[2]{\langle #1 \rvert #2 \rangle}
\newcommand{\ketbra}[2]{\left| {#1} \right\rangle\left\langle {#2}\right|}
\newcommand{\ud}{\mathrm{d}}
\newcommand{\mmax}{_{\text{max}}}
\newcommand{\thetabar}{\overline{\theta}}
\begin{document}

\title{Bell inequalities with continuous angular variables}

\author{Carolina V. S. Borges$^{1,2,3}$, P\'{e}rola Milman$^{1,2}$ and Arne Keller$^1$}

\affiliation{$^{1}$
Univ. Paris-Sud 11, Institut de Sciences Mol\'{e}culaires d'Orsay (CNRS),
B\^{a}timent 210--Campus d'Orsay, 91405 Orsay Cedex, France}
\affiliation{$^{2}$
Laboratoire Mat\'{e}riaux et Ph\'{e}nom\`{e}nes Quantiques, Universit\'{e} Paris Diderot, CNRS UMR 7162, Universit\'{e} Paris Diderot, 75013, Paris, France.}
\affiliation{$^{3}$
Instituto de F\'{\i}sica, Universidade Federal Fluminense, Niter\'{o}i, RJ, Brazil
}

\date{\today}
\begin{abstract}
We consider  bipartite quantum systems characterized by a continuous angular variable $\theta \in [-\pi, \pi[$, representing, for instance, the position of a particle on a circle. We show how to reveal non-locality on this type of system using inequalities similar to CHSH ones, originally derived for bipartite spin $1/2$ like systems. Such inequalities involve correlated measurement of continuous angular functions and are equivalent to the continuous superposition of CHSH inequalities acting on bidimensional subspaces of the infinite dimensional Hilbert space. As an example, we discuss in detail one application of our results, and we derive inequalities based on orientation correlation measurements.  The introduced Bell-type inequalities open the perspective of new and simpler experiments to test non locality on a variety of quantum systems described by continuous variables.
\end{abstract}

\pacs{03.65.Ud;03.67.-a;33.20.Sn}

\maketitle

{\it Introduction:} The pioneering discussions of Einstein, Podolsky and Rosen (EPR) \cite{EPR}, which rose the possibility of eventual conflict between the classical an the quantum definitions of realism and locality, dealt with measurements of continuous variables (CV) of a quantum system, as  position $q$ and  the kinetic momentum $p$. Almost $30$ years elapsed before J. S. Bell promoted a regain of interest on the subject, deriving a quantitative criteria establishing the border between quantum and classical physics concerning locality and realism \cite{Bell1964}. The discretized version of the EPR paradox and of the Bell inequalities using spin-$1/2$ like systems came with the work of Bohm \cite{Bohm1957} and Clauser, Horne, Shimony and Holt (CHSH) \cite{CHSH1969} and was motivated by the necessity to find simpler setups for experimentally testing the, up to then, Gedankenexperiments exposing the quantum-classical contradiction \cite{Aspect1982}. The almost 30 years following such pioneering experiments led to theoretical and experimental advances pointing out that using CV may provide advantages with respect to discrete systems. Higher intensity signals allowing to circumvent the detection loophole are among the announced advantages \cite{Gilchrist1998,Cavalcanti2007}. Most of these works are based on canonical unbounded variables, as $(q,p)$. They can correspond, for instance, to the sum and the difference of two quadratures of the electromagnetic field. The first attempts to build CV non-locality tests consisted on finding ways to discretize  the continuum to reuse concepts  developed for the discrete case. It was shown in \cite{Banaszek1999} that dichotomizing the phase space according to a state's parity and its displacement in phase space can lead to Bell type inequalities that can be violated by gaussian continuous variable entangled states. Other phase space dichotomizations are possible, as the one proposed in \cite{Wenger2003}. However, dichotomization  does not correspond to a genuine CV measurement, since {\it it provides an observable with a discrete, rather than genuinely continuous spectrum}.

Reid and co-workers \cite{Cavalcanti2007} proposed variance based Bell-type inequalities not demanding dichotomization, that are independent of the spectrum of the measured observables. Another possible approach are entropy based inequalities that may be advantageous to detect entanglement in non-gaussian states, since they go beyond the usual criteria that use second-order moments \cite{Walborn2009,Rojas1995}.

\begin{figure}
\begin{center}
\includegraphics[scale=.2]{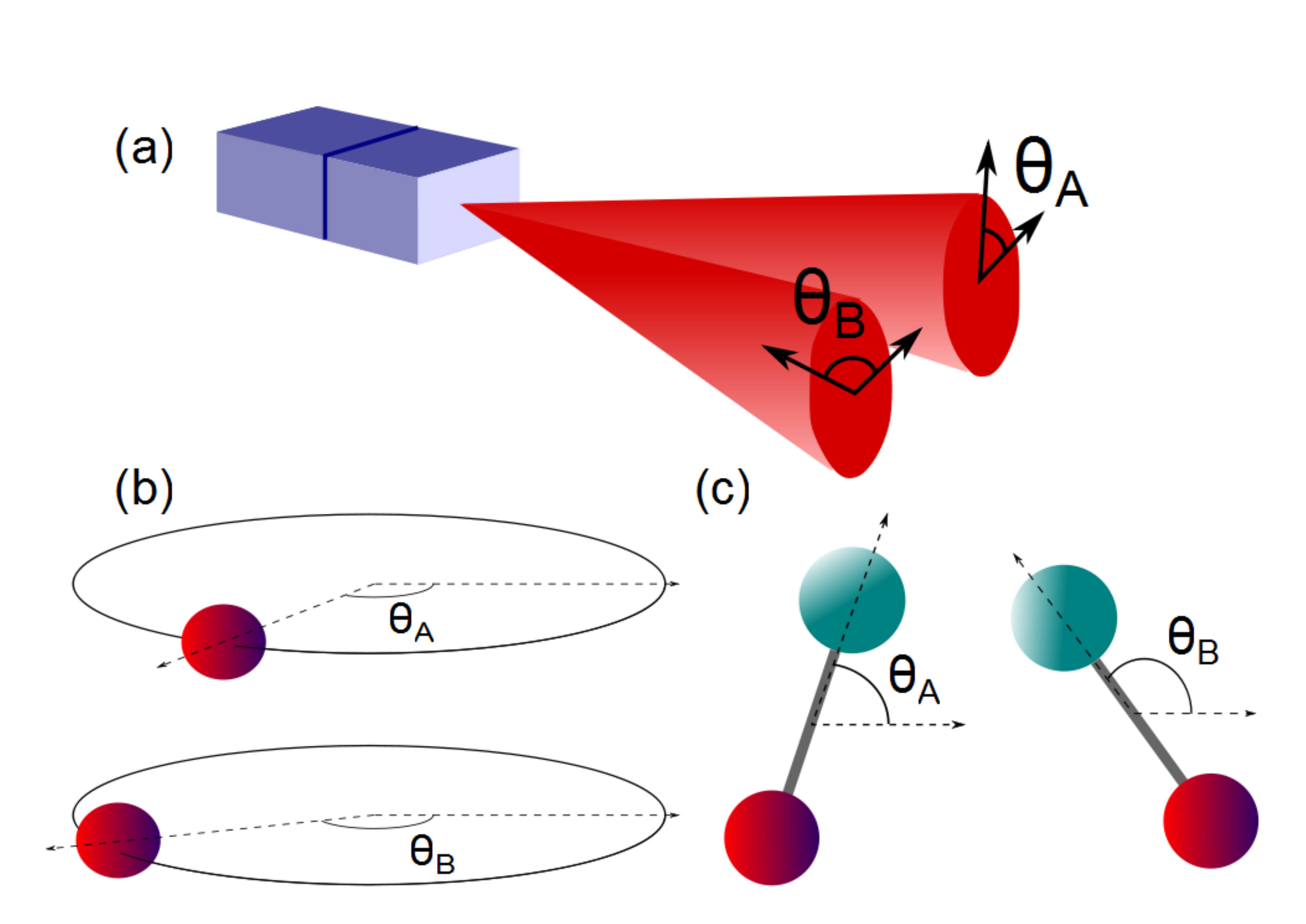}
\end{center}
\caption{Examples of physical systems for which our results can be applied: (a) The transverse profile of a propagating light beam (b) Particles moving on a circle (as charged particles in an electromagnetic trap) (c) rigid rotators confined to two dimensional motion.   \label{fig1}}
\end{figure}

Up to now, most of the results on CV entanglement and non locality have been devoted to the case of canonical variables with an unbounded spectrum, as position and momentum. In the present work, we consider a different type of CV. We deal here with two quantum systems, $A$ and $B$, characterized by angular variables $\theta_i\in[-\pi,\pi[$, $i=A,B$ on a circle, instead of a position $q$ in the full real line. $\theta$ can represent, for instance,  the position of a particle on a circle, a rotator on a plane (for example, molecules confined to a plane \cite{Svensson1999}), or the polar angle locating the photon field in the plane transverse to its propagation, as illustrated in Fig. \ref{fig1}. The main result of the present paper is to build  inequalities involving correlated measurement of continuous, real and periodic angular functions $f_i(\theta_i)$  ($i=A,B$). Surprisingly, the derived inequalities share similarities to CHSH-like ones. As a matter of fact,  their general form is given by a properly balanced continuous superposition of CHSH-type inequalities. Exploiting the possibility of revealing non locality using angular measurements is an approach that can be particularly useful and suitable for a number of physical systems in atomic and  molecular physics, photonics and mesoscopic systems in condensed matter physics.

{\it Theoretical model: } As a starting point, we recall  the CHSH inequalities for a bipartite two-level system. It can be written in a simple form as  $\abs{\moy{B(a,a',b,b')}}\leq2$, with:
\beq
\label{eq:CHSH}
B(a,a',b,b') = \sigma_a\otimes\sigma_b + \sigma_a\otimes\sigma_{b'} + \sigma_{a'}\otimes\sigma_b - \sigma_{a'}\otimes\sigma_{b'}
\eeq
where $\sigma_{\alpha}$ is the Pauli matrix in one direction $\alpha$  of the tridimensional space. $a$ and ${a'}$ refer to  different directions of the spin projection of subsystem $A$ while $b$ and ${b'}$ are associated to subsystem $B$. The inequality $\abs{\moy{B(a,a',b,b')}}\leq2$ is fulfilled in the framework of local hidden variable (LHV) theories. The fact that there exist quantum states violating these inequalities disproof quantum mechanics as a  local or EPR realistic theory. For quantum mechanical systems, \eqref{eq:CHSH} is bounded by the Tsirelson bound of $2\sqrt{2}$ \cite{Tsirelson1980}.

We now consider a different situation of measurement providing continuous and $2\pi$-periodic outcomes. In other words, the underlying Hilbert space is the square integrable  $2\pi$-periodic functions space ($L^2([-\pi,\pi[,d\theta)$). Such functions can be spanned by the basis  $\left\{\ket{m_i}; m_i\in\mathbb{Z}\right\}$, which are angular momentum eigenstates: $J_z\ket{m_i}=m\ket{m_i}$. An alternative continuous basis is $\left\{\ket{\theta_i};\theta_i\in[-\pi,\pi[\right\}$, and it can be obtained from the $\ket{m_i}$ basis by Fourier transform: $\braket{m_i}{\theta_i}=\frac{1}{\sqrt{2\pi}}e^{im_i\theta_i}$. In this representation, a local observable for each party can be written as  $F_i=\int_{-\pi}^{\pi}d\theta_if_i(\theta_i)\ket{\theta_i}\bra{\theta_i}$, with $f_i(\theta_i)$ real, bounded and periodic.

Our guidelines to obtain continuous variables Bell inequalities is to build a CHSH~\cite{CHSH1969} Bell--operator similar to the one given by \eqref{eq:CHSH}, but based on the correlated measurements of $F_i$  for each particle. It is clear that, under the assumption that the spectra of $F_i$ are bounded, this property is preserved by every unitary transformation $U_i(\phi_i)$ such that $F(\phi_i)=U_i(\phi_i)F_iU_i^{\dagger}(\phi_i)$. It is also straightforward to show that, for a LHV theory, we have:
\bea \label{BellCont}
&&\abs{\moy{\mathcal{B}(\phi_a, \phi_a',\phi_b,\phi_b')}} =  \nonumber \\
&& \left|\left\langle F_A(\phi_a)\otimes F_B(\phi_b) +F_A(\phi_a')\otimes F_B(\phi_b) \right.\right. \nonumber \\
&&\left.\left.+F_A(\phi_a)\otimes F_B(\phi_b') -F_A(\phi_a')\otimes F_B(\phi_b')\right\rangle\right| \leq 2,
\eea
if the maximum value of $f_i(\theta_i)$ is normalized to $1$. What are the conditions the functions $f_i(\theta_i)$ and the transformed operators $F_i(\phi_i)$ should satisfy to violate (\ref{BellCont}) and allow for non-locality and entanglement tests? In order to answer this question, we focus on $f_i(\theta_i)=\cos(\theta_i)$ which is a function that can be measured  on a variety of physical systems. The results derived in the following can be straightforwardly generalized  to all $2\pi$ periodic functions such that $f(\theta)=-f(\theta-\pi)$ $\forall\theta\in[0,\pi]$.

{\it Example of application: } The observables $C=F_i$ corresponding to  $f_i(\theta_i)=\cos(\theta_i)$,  can be expressed on the $\ket{m}$ basis as:
\beq
C=\sum_{m\in\mathbb{Z}}\frac{1}{2}\left (\ket{m+1}\bra{m}+\ket{m}\bra{m+1}\right ).
\eeq
For a particle rotating on a circle, this operator corresponds to the projection on the $x$ axis of  the particle position; it can also be related to the spatial orientation of a two dimensional rotor, or to the phase of a superconducting circuit. Operator $C$ has a spectrum of doubly degenerated eigenstates $\ket{\theta}$ and $\ket{-\theta}$, both with eigenvalue $\cos\theta$. For this operator, we can define an equivalent to the rotation of a spin $1/2$ system: it is the unitary operator $e^{iJ_z^2\phi}$, (that is, the free evolution operator during a time $t=2\hbar\phi$, for a free particle with unit mass and angular momentum $J_z$, constrained to move on an unit circle).
From $C$, we can define $C(\phi_i)\equiv F_i(\phi_i)$ as:
\bea
\label{eq:Cphi}
C(\phi) =e^{i J_{z}^2\phi} Ce^{-i J_{z}^2\phi} &=&\frac{1}{2}\sum_{m \in \mathbb{Z}} e^{i(2m+1)\phi}\ket{m}\bra{m+1} +\nonumber \\
& &e^{-i(2m+1)\phi}\ket{m+1}\bra{m},
\eea
and for a bipartite system (particles $A$ and $B$) the Bell operator (\ref{BellCont}) reads: $\mathcal{B}(\phi_a,\phi_a',\phi_b,\phi_b')=C_A(\phi_a)\otimes C_B(\phi_b)+C_A(\phi_a')\otimes C_B(\phi_b)+C_A(\phi_a)\otimes C_B(\phi_b')-C_A(\phi_a')\otimes C_B(\phi_b')$,
where $C_{A(B)}(\phi)$ are operators defined as in Eq.~\eqref{eq:Cphi} acting on the Hilbert space of particle $A(B)$. The spectrum of $C(\phi)$ does not depend on $\phi$. Diagonalizing $C(\phi)$ shows that its spectrum is bounded,  with $\abs{\moy{C}}\leq1$. Thus, for a LHV theory, $\abs{\moy{\mathcal{B}(\phi_a,\phi_a',\phi_b,\phi_b')}}<2$ holds. However,  for some set of phases $\phi_i$s and some quantum states, this  inequality is violated. In order to show that, we calculate the spectra of $\mathcal{B}(\phi_a,\phi_a',\phi_b,\phi_b')$ operators. Actually, such spectra depend only on the relative phases $\phi_a-\phi_a'$ and $\phi_b-\phi_b'$  \cite{moleculesEPJD,moleculesPRL}. We can define $B(\xi_a,\xi_b)=B(\phi_a'-\phi_a,\phi_b'-\phi_b)\equiv\mathcal{B}(0,\phi_a'-\phi_a,0,\phi_b'-\phi_b)=e^{iJ^2_{Az}\phi_a}\otimes e^{iJ^2_{Bz}\phi_b}\mathcal{B}(\phi_a,\phi_a',\phi_b,\phi_b')e^{-iJ^2_{Az}\phi_a}\otimes e^{-iJ^2_{Bz}\phi_b}$. $B(\xi_a,\xi_b)$ and $\mathcal{B}(\phi_a,\phi_a',\phi_b,\phi_b')$ are related by an unitary transformation. Therefore, the variation of only the two phases $(\xi_a,\xi_b)$ is enough to explore the spectrum of all the $\mathcal{B}(\phi_a,\phi_a',\phi_b,\phi_b')$.

\begin{figure}
\begin{center}
\includegraphics[scale=.3]{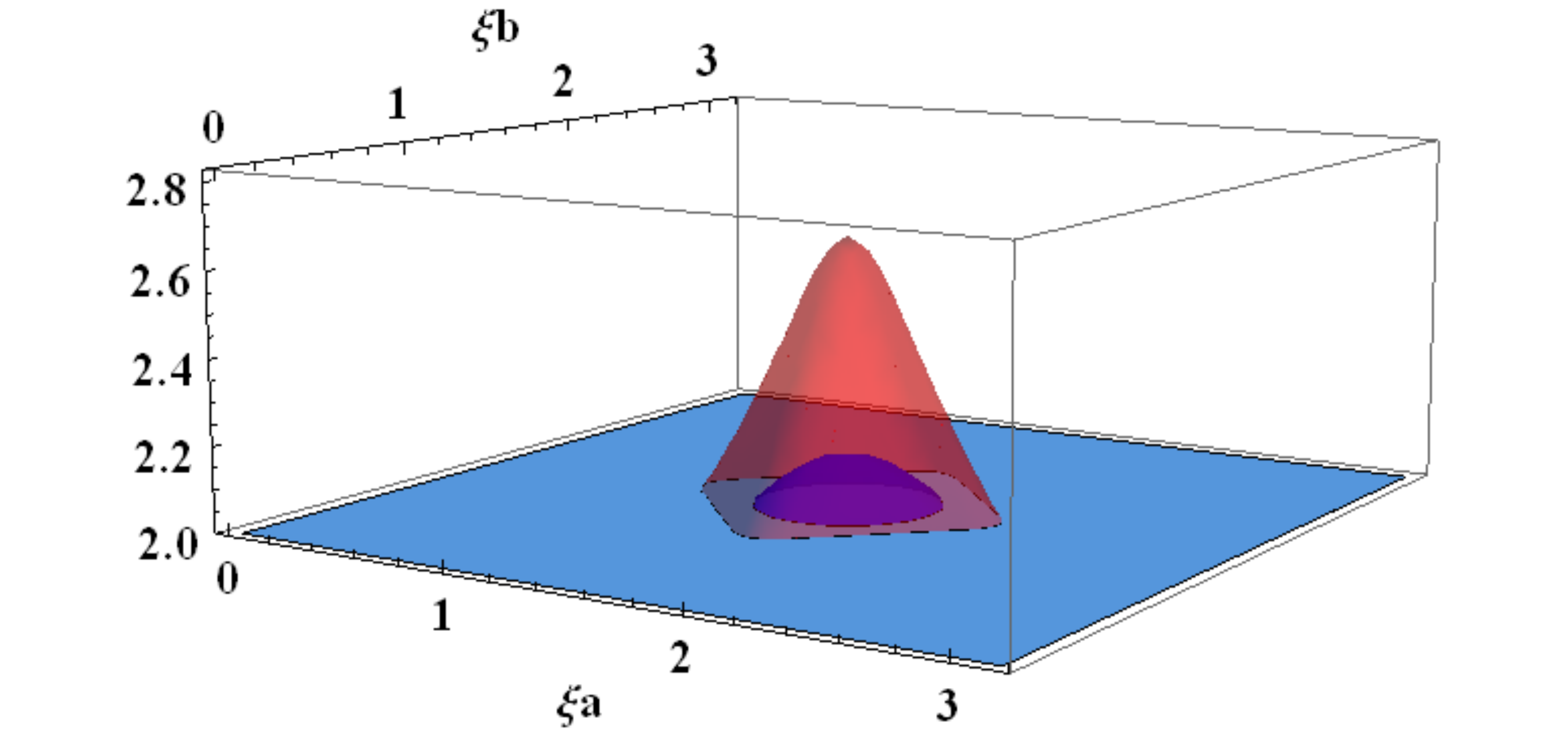}
\end{center}
\caption{Plot of the maximum eigenvalue of $B^{(M)}(\xi_a,\xi_b)$, $b^{(M)}\mmax$, in the region of violation ($b^{(M)}\mmax>2$) as a function of $\xi_a$ and $\xi_b$ for $M=2$ (inner blue plot), and $M=5$ (outer red plot). \label{fig2}}
\end{figure}

As a first approach to the study of the dependence of operators $B(\xi_{a},\xi_{b})$ with the phases $\xi_i$ , we discretize the possible outcomes of the  correlation measurements. For this purpose we consider that the measured quantum states  lie on a $2M+1$ finite dimension space generated by the basis set $\left\{\ket{m};m\in\mathbb{Z},\abs{m}\leq M\right\}$.
We thus define  the projection $C^{(M)}(\xi)$  of the $C(\xi)$ operator on this reduced space as~:
\beq
\label{opcosM}
C^{(M)} (\xi)=\frac{1}{2}\sum_{\abs{m}\leq M} e^{i(2m+1)\xi}\ket{m}\bra{m+1}
+e^{-i(2m+1)\xi}\ket{m+1}\bra{m}
\eeq
and $C^{(M)}=C^{(M)}(0)$.
We also define  discretized Bell operators $\mathcal{B}^{(M)}$ and $B^{(M)}$,  analogously to the continuous case, but where the $C(\xi_i)$ operators are replaced by $C^{(M)}(\xi_i)$ operators, which act on the considered finite dimensional space only.  Discretization allows to explicitly compute the $2M+1$ eigenstates and corresponding eigenvalues of $C^{(M)}$, providing thus a more intuitive physical image of the considered operators. Also, it enables the numerical study of the spectrum of $B^{(M)}(\xi_a,\xi_b)$ and its phase dependency. Defining  $b^{(M)}_{\text{max}}(\xi_a,\xi_a)$ as the highest eigenvalue of $B^{(M)}(\xi_a,\xi_b)$ we see in Fig. \ref{fig2} that, for $M=2$ and $M=5$, $b^{(M)}_{\text{max}}(\xi_a,\xi_a)$ reaches its maximum at $\xi_a=\xi_b=\pi/2$. We verified numerically that this fact is independent of $M$ for $1\leq M\leq20$. In addition, $|b(\pi/2,\pi/2)|$ increases as $M$ increases, being already greater than $2$ (i.e., enabling the Bell-type inequality violation) for $M=2$.

The study of the discretized operators is now used as a starting point to the derivation of the main results of this Letter, which are non-locality and entanglement tests in the continuous limit. We start by analyzing the point $\xi_a=\xi_b=\pi/2$, since it is the one that  provides the highest contrast between $b\mmax(\xi_a,\xi_a)$ and the violation threshold value ($2$).

We then move to the continuous limit  ($M\rightarrow\infty$) and search for the highest possible value of violation  and the corresponding non local states for this choice of phases. Using the expression of $C(\xi)$ \eqref{eq:Cphi} and the anti-symmetric property of $\cos{\theta}$, we compute
$\bra{\theta'}C(\frac{\pi}{2})\ket{\theta}$ explicitly:
\bea
\label{eq:Cpideux}
C(\frac{\pi}{2}) =
\int_{0}^{\pi}
\ud\theta
\cos\theta
\left[
i\ket{\theta-\pi}\bra{\theta}  -
i \ket{\theta}\bra{\theta-\pi}
\right].
\eea
It is now convenient to define the following operators~:
\begin{eqnarray}
\label{eq:contpaulix}
\sigma^x(\theta) &= \ket{\theta}\bra{\thetabar}  + \ket{\thetabar}\bra{\theta}\nonumber \\
\sigma^y(\theta) &= i\left(\ket{\thetabar}\bra{\theta}  -\ket{\theta}\bra{\thetabar} \right)\\
\sigma^z(\theta) &= \ket{\theta}\bra{\theta}  - \ket{\thetabar}\bra{\thetabar},\nonumber
\end{eqnarray}
for $\theta\in[0,\pi[$ and with $\thetabar\equiv\theta-\pi$. These operators constitute an orthogonal continuous set of Pauli-like operators, and they fulfill relations analog to
the usual Pauli matrices: $\sigma^j(\theta)\sigma^k(\theta')=i\sigma^l(\theta)\delta(\theta-\theta')$ where $(j,k,l)$ denotes circular permutations of $(x,y,z)$. They also fulfill orthogonality relations $\sigma^j(\theta)\sigma^j(\theta')=\delta(\theta-\theta')(\ketbra{\theta}{\theta}+\ketbra{\thetabar}{\thetabar})$.
Eq.\eqref{eq:Cpideux} can thus be recast as:
\bea
\label{eq:Cdiag}
C &=& \int_{0}^{\pi} \ud\theta \cos\theta \,\sigma^z(\theta) \\
C(\frac{\pi}{2}) &=& -\int_{0}^{\pi} \ud\theta \cos\theta \sigma^y(\theta).
\eea
Thus, the Bell operator $B_{\text{m}}\equiv B(\frac{\pi}{2},\frac{\pi}{2})$ can  be written as the following direct sum:
\begin{align}
\label{eq:Bsigma}
B_{\text{m}} =
  \int_{0}^{\pi}\int_{0}^{\pi}\ud\theta \ud\theta' \cos\theta \cos\theta' X(\theta,\theta')
\end{align}
where
\begin{align}
\label{eq:X}
&X(\theta,\theta') = \left[\sigma^{z}_{A}(\theta)\otimes \sigma^{z}_{B}(\theta') - \sigma^{y}_{A}(\theta)\otimes
\sigma^{z}_{B}(\theta')  \right. \nonumber\\
 &-\left. \sigma^{z}_{A}(\theta)\otimes \sigma^{y}_{B}(\theta') - \sigma^{y}_{A}(\theta)\otimes \sigma^{y}_{B }(\theta')
\right],
\end{align}
and $\sigma^{j}_{A}(\theta)$ and $\sigma^{j}_{B}(\theta)$  $(j=x,y,z)$ are operators defined by Eqs.~\eqref{eq:contpaulix} on the Hilbert space of parties $A$ and $B$ respectively. The $X(\theta,\theta')$ operators are orthogonal:
\begin{align}
\label{eq:Xortho}
&X(\theta_a,\theta_b) X(\theta'_a,\theta'_b) =
4  \delta(\theta_a-\theta_a') \delta(\theta_b-\theta_b') \times \nonumber \\
 &(\ketbra{\theta_a}{\theta_a}  + \ketbra{\overline{\theta_a}}{\overline{\theta_a}}) \otimes
(\ketbra{\theta_b}{\theta_b}  + \ketbra{\overline{\theta_b}}{\overline{\theta_b}})
\end{align}
and completely analog to the usual 2--qubits (4-dimensional)  CHSH operators.
We have thus the surprising result that the Bell operator $B_{\text{m}}$ is the weighted continuous direct sum of  2--qubit-like CHSH Bell operators $X(\theta,\theta')$, with the weights being given by $\cos\theta\cos\theta'$.
Thanks to the orthogonality property \eqref{eq:Xortho}, finding the spectrum and eigenstates of $B_{\text{m}}$ is a simple task. Indeed, for each $\theta$ and $\theta'$ it is enough to diagonalize the $4\times4$ matrix representing $X(\theta,\theta')$.
We find that $B_{\text{m}}$ can thus be written in diagonal form as:
\bea
B_{\text{m}}&=&2\sqrt{2}\sum_{n=\pm1}n\int_{0}^{\pi}\int_{0}^{\pi}\ud\theta \ud\theta' \times\nonumber\\
&&\cos\theta\cos\theta'\ketbra{\chi^n(\theta,\theta')}{\chi^n(\theta,\theta')}
\eea
where~:
\begin{align}\label{chi}
&\ket{\chi^{\pm 1 }(\theta,\theta')} = \frac{1}{N_{\pm}} [\ket{\theta}\otimes\ket{\theta'} + \ket{\thetabar}\otimes\ket{\overline{\theta'}} \nonumber\\
&\mp i(\sqrt{2}\mp1)\left(\ket{\theta}\otimes\ket{\overline{\theta'}} + \ket{\thetabar}\otimes\ket{\theta'}\right)]
\end{align}
are the eigenvectors of $X(\theta,\theta')$ and of $B_{\text{m}}$ with non-zero eigenvalues and
where $N_{\pm}=\left[2(2\mp\sqrt{2})\right]^{1/2}$ is a normalization factor such that~:
$\braket{\chi^{n'}(\theta_a',\theta_b')}{\chi^{n}(\theta_a,\theta_b)}=\delta_{nn'}\delta(\theta_a'-\theta_a)\delta(\theta_b'-\theta_b)$.

Therefore, the spectrum of $B_\text{m}$ is continuous and equal to $[-2\sqrt{2},2\sqrt{2}]$. $\abs{\moy{B_\text{m}}}$ is thus bounded by $2\sqrt{2}$ as in the 2-qubit CHSH inequality. However, the values $\pm2\sqrt{2}$ are only reached by eigenstates of $B_\text{m}$ which are not physical states since they are composed of perfectly oriented states, namely $\ket{0}$ and $\ket{\pi}$. In the angular representation, these states are expressed by delta functions, since $\braket{\theta}{\theta'}=\delta(\theta-\theta')$.

In order to explicitly construct physically sound states violating our Bell inequality, we should consider wave packets consisting of continuous superposition of eigenstates $\ket{\chi^{+1}(\theta,\theta')}$, with $\theta$ and $\theta'$  localized around
$\theta=\theta'=0$, point of the maximum eigenvalue of $B_{\text{m}}$. An example of such wave packet is given by:
\beq\label{pack}
\ket{\Psi} = \int_0^{\pi} \ud\theta \int_0^{\pi}  \ud\theta' g_a(\theta)g_b(\theta') \ket{\chi^{+1}(\theta,\theta')}.
\eeq
where $g_a(\theta)$ and $g_b(\theta)$ are normalized $L^2(]0,\pi[,d\theta)$ functions with support containing $\theta=0$.
The expectation value of ${B}_{\text{m}}$ for this state is
\beq
\label{eq:moyBm}
\moy{B_{\text{m}}}_{\Psi} = 2\sqrt{2}\int_0^{\pi} \ud\theta \int_0^{\pi} \ud\theta' \cos\theta \cos\theta' |g_a(\theta)|^2|g_b(\theta')|^2
\eeq
The wave packet (\ref{pack}) can be produced making a linear combination of the one--particle wave packets
\beq
\label{eq:g}
\ket{g}=\int_{0}^{\pi} \ud\theta g(\theta)\ket{\theta} \text{ and }
\ket{\overline{g}}=\int_{0}^{\pi} \ud\theta g(\theta)\ket{\thetabar},
\eeq
taking  the same coefficients of Eq.~(\ref{chi}):
\bea\label{Psi}
\ket{\Psi}&=& \frac{1}{\sqrt{2} N_{+}} [(\ket{g_a}\otimes\ket{g_b} + \ket{\overline{g_a}}\otimes\ket{\overline{g_b}}) \\
&-& i(\sqrt{2}-1)\left(\ket{g_a}\otimes\ket{\overline{g_b}} + \ket{\overline{g_a}}\otimes\ket{g_b}\right)].\nonumber
\eea
It is useful now to obtain a simple relation between the wave packet localization and the violation of the Bell inequality. For such, we
take for $g(\theta)$ the ideal case of an angular slit with aperture $\delta\theta$,  given by~:
\beq
\label{eq:angslit}
g_a(\theta)= g_b(\theta)=\left\{ \begin{array}{cc}
\frac{1}{\sqrt{\delta \theta}} & \text{for } \theta < \delta \theta \\
0 & \text{otherwise}  \end{array} \right.
\eeq
Using Eq.~\eqref{eq:moyBm}, the value of $\moy{B_{\text{m}}}_{\Psi}$ can be written as function of the aperture $\delta\theta$ of the slit as follows:
\beq
\moy{B_{\text{m}}}_{\Psi}=2\sqrt{2}\left(\frac{\sin\delta \theta}{\delta \theta}\right)^2.
\eeq
This equation above shows that we can obtain values of $\moy{B_{\text{m}}}_{\Psi}>2$, violating the Bell inequality, with an aperture of $\delta\theta<18^{\circ}$ which is not a too restrictive condition. There is thus a relatively broad collection of simple two--particles non local pure states involving coherent superposition of
one--particle wave packets localized around $\theta=0$ and $\theta=\pi$ that violate the derived Bell inequality.

We now study the example of even more realistic states, which are non-pure ones, establishing some conditions for them to violate the derived Bell-type inequalities. We consider the analog of the Werner states \cite{Werner1989} : $\rho(\eta)=\eta\rho_A\otimes\rho_B+(1-\eta)\ketbra{\Psi}{\Psi}$ \cite{Mista2002}, where $\rho_{A(B)}=\text{Tr}_{B(A)}\left[\ketbra{\Psi}{\Psi}\right]=\frac{1}{2}\left(\ketbra{g_{a(b)}}{g_{a(b)}}+\ketbra{\overline{g_{a(b)}}}{\overline{g_{a(b)}}}\right)$, with $\ket{\Psi}$ given by \eqref{Psi}. As $\bra{\chi^n(\theta_a,\theta_b)}\rho_A\otimes\rho_B\ket{\chi^n(\theta_a,\theta_b)}$ does not depend on $n=\pm1$ for all $\theta_a,\theta_b\in[0,\pi]$, then
$\text{Tr}\left[B_\text{m}\rho_A\otimes\rho_B\right]=0$. It turns out that  the expectation of $B_\text{m}$ for the state $\rho(\eta)$ is simply given by $(1-\eta)\moy{B_\text{m}}_{\Psi}$. In the ideal case where $g_a$ and $g_b$ are given by \eqref{eq:angslit}, the maximal allowed value of the mixing coefficient $\eta$ for the Bell equation to be violated is a simple function of the slit aperture~: $\text{Tr}\left[B_\text{m}\rho(\eta)\right]>2\Rightarrow\eta<1-\frac{1}{\sqrt{2}}\left(\frac{\delta\theta}{\sin\delta\theta}\right)^2$, providing conditions for this type of mixed states to violate non-locality.
{\it Conclusion:}  we derived general CHSH-like Bell-type inequalities for continuous variables using measurements of bounded observables with anti-symmetric and periodic spectra. Such observables are relevant for a number of experimental systems, ranging from the photonic orbital angular momentum to the movement of material particles on a circle and the phase of  superconducting currents. We discussed in detail an example of observable, but the results obtained are valid for {\it all observables with the same spectral properties and symmetries}.  Our results exploit angular variables from an original approach, and open the path to novel experiments with a wide range of quantum devices, throwing some light on the difficult task of non-locality violation with genuinely continuous variables.

Authors acknowledge funding by CAPES-COFECUB No. 640/09.

\end{document}